\documentclass[aps,prd,twocolumn,showpacs,groupedaddress]{revtex4}
\usepackage{epsfig}
\usepackage{longtable}
\usepackage{amsfonts}

\usepackage{amsmath}
\usepackage{amsfonts}
\usepackage{amssymb}
\usepackage{graphicx}
\newcommand{\be}{\begin{equation}}
\newcommand{\ee}{\end{equation}}

\begin{document}
%\draft
%\preprint{HEP/123-qed}

\title{Temperature measurements of a Bose--Einstein condensate by ultra--intense light pulses}

\author{Abel Camacho}
\email{acq@xanum.uam.mx} \affiliation{Departamento de F\'{\i}sica,
Universidad Aut\'onoma Metropolitana--Iztapalapa\\
Apartado Postal 55--534, C.P. 09340, M\'exico, D.F., M\'exico.}%Lines break automatically or can be forced with \\

\author{Luis F. Barrag\'{a}n-Gil}
\email{brgl@xanum.uam.mx} \affiliation{Departamento de
F\'{\i}sica, Universidad Aut\'onoma Metropolitana--Iztapalapa\\
Apartado Postal 55--534, C.P. 09340, M\'exico, D.F., M\'exico.}

\author{Alfredo Mac\'{\i}as}
\email{amac@xanum.uam.mx} \affiliation{Departamento de
F\'{\i}sica,
Universidad Aut\'onoma Metropolitana--Iztapalapa\\
Apartado Postal 55--534, C.P. 09340, M\'exico, D.F., M\'exico.}

\date{\today}% It is always \today, today,

\begin{abstract}
Experimentally the temperature in a Bose--Einstein condensate is
always deduced resorting to the comparison between the
Maxwell--Boltzmann velocity distribution function and the density
profile in momentum space. Though a successful method it is an
approximation, since it also implies the use of classical
statistical mechanics at temperatures close to the condensation
temperature where quantal effects play a relevant role. The
present work puts forward a new method in which we use an
ultra--intense light pulse and a nonlinear optical material as
detectors for differences in times--of--flight. This experimental
value shall be compared against the result here calculated, using
the Bose--Einstein distribution function, which is a
temperature--dependent variable, and in this way the temperature
of the condensate is obtained.
\end{abstract}


\maketitle
%*****************************************
%\section{Introduction}

The emergence of the concept of Bose--Einstein condensate (BEC)
dates back to the year of 1924 with a paper by S. N. Bose in which
the statistics of the quanta of light was analyzed \cite{BE}. A
year later A. Einstein with the idea of Bose predicted the
occurrence of a phase transition in a gas of non--interacting
atoms \cite{Einstein}. It can be stated that a BEC is a state of
bosonic matter, confined by an external potential, and cooled to
temperatures very close to the absolute zero \cite{Stenholm}.
Under these conditions the atoms begin to collapse into the lowest
energy wave function associated to the corresponding confinement
potential \cite{Phatria}.

Since our proposal involves the use of the features of nonlinear
optics let us explain, very briefly, the main idea behind this
topic, which emerged in 1961 \cite{Franken}. As a light beam
propagates through a piece of material the polarization of its
atoms emerges as a response to the excitation defined by the
corresponding electric field. As long as the intensity of the
electric field remains smaller than a certain threshold, the
interaction is linear and the transmitted light has the same
frequency as the incident one \cite{Franken}. But if the intensity
of the light is beyond this threshold, then the transmitted light
has a spectrum with more than one component, among which the
double frequency is often the most important \cite{Brabec}.

%********************************************************
%\section{Maxwell--Boltzmann Distribution and the temperature in a BEC}

The experimental achievement of BEC has spurred the appearance of
many questions in the theoretical realm \cite{Lev} as well as in
the experimental level. One of them is related to the procedure by
which the temperature is deduced. Indeed, the temperature is
always obtained resorting to an absorption imaging method
\cite{Davis1}. In this procedure the corresponding atom cloud is
imaged either while it was trapped or following a switch--off of
the used trap. The time--of--flight images of the atoms are
compared against a Maxwell--Boltzmann velocity distribution
function \cite{Phatria}, and by this way the temperature is
deduced. It is noteworthy to mention that in the context of this
issue the deduction of the temperature is always done resorting to
a statistics, Maxwell--Boltzmann, which is valid in the regime of
high temperatures, i.e., far away from the condensation
temperature \cite{Phatria}. It can be argued that it is a good
approximation (see \cite{Pethick} pp. 27 and \cite{Dalfovo} pp.
469). Nevertheless,  from a conceptual point of view this
procedure has a severe shortcoming since it resorts to a model
(Maxwell--Boltzmann statistics) in a region in which it loses its
validity. This last remark raises the following question: could
the temperature of a BEC be deduced without the use of
Maxwell--Boltzmann statistics?

The present work provides an affirmative answer to this question.
Our proposal involves an ultra--intense light pulse. This pulse
will be divided into two beams, one of them will travel through
the condensate while the other one will not (see Fig.(\ref{f-1})).
The difference in time--of--flight between these two beams will be
detected resorting to a nonlinear optical material. Additionally,
we deduce the expression for the difference in time--of--flight as
a function of the parameters of the trap, and hence we obtain an
expression for the temperature. It has to be stressed that in this
process classical statistics is not used. Since the required laser
fields generate ultrashort light pulses we may detect phenomena
involving times scales in the range of femtoseconds \cite{Brabec}.
This means that our procedure offers us the possibility of
detecting differences in times--of--flight around this order of
magnitude.

%*********************************************
%\section{Light Interferometry and temperature}

%****************************************
%\subsection{Refraction index and trapping parameters}

Consider a laser light beam, with wavelength $\lambda$, and divide
it into two parts. One of them will be used as a reference beam,
and the second one will pass through the corresponding condensate.
Afterwards, both beams are brought together and then the
difference in time--of--flight between them is calculated.

Let us now proceed to obtain the time--of--flight as function of
the variables of the condensate. The density of the gas
$n(\vec{r})$ can be related to the refraction index
$\tilde{n}(\vec{r})$ and the particle size $R$ \cite{Newton} as
follows (a condition to be satisfied is $\lambda^2Rn(\vec{r})<1$)
\begin{eqnarray}
 \tilde{n}(\vec{r})= 1 + \lambda^2Rn(\vec{r})
 \label{refraction1} .
\end{eqnarray}

The density, which is connected to the eigenfunctions of the
trapped particles $\phi_{\nu}(\vec{r})$ and the mean occupation
number $\langle n_{\nu}\rangle$ \cite{Pethick}, reads
\begin{eqnarray}
 n(\vec{r}) = \sum_{\nu}\Bigl[\langle n_{\nu}\rangle | \phi_{\nu}(\vec{r})|^{2}\Bigr] \label{density1} .
\end{eqnarray}

For our case, a Bose--Einstein situation, the occupation number is
a function of the energy eigenvalues $\epsilon_{\nu}$, the
temperature, and the chemical potential $\mu$ \cite{Phatria},
i.e.,
\begin{eqnarray}
 \langle n_{\nu}\rangle =  \frac{1}{\exp{\Bigl[(\epsilon_{\nu} - \mu)/kT\Bigr]} -1} \label{occnumber} .
\end{eqnarray}

These last three expressions show that the refraction index will
depend upon the temperature, and, in consequence, the difference
in time--of--flight will be a function of $T$. This is the point
to be exploited in the present work. We now consider the case of a
Bose--Einstein gas of non--interacting and non--relativistic
particles immersed in a Newtonian homogeneous gravitational field.
The inclusion of gravity has been done because, at least for
alkali atoms, it has non--negligible effects \cite{Legget,
Remington}. Additionally, an anisotropic harmonic--oscillator
confining potential is present. At this point we must add that in
the usual experimental devices the atom clouds are confined with
the help of laser trapping or magnetic traps \cite{Legget}. In the
case of alkali atoms some of the available confining potentials
can be approximated by a three--dimensional harmonic oscillator
(see \cite{Dalfovo} pp. 466). In other words, our potential
represents a very good approximation to some experimental cases.

The frequencies of our harmonic oscillators along the
coordinate--axes will be denoted by $\omega_l$, where $l= x, y,
z$. In addition, $g$ represents the acceleration of gravity. Under
these conditions the complete potential reads
\begin{eqnarray}
V(x, y, z) = \frac{m}{2}\Bigl[\omega_x^2x^2 + \omega_y^2y^2 +
\omega_z^2z^2\Bigr]  + mgz \label{potential1}.
\end{eqnarray}

In general the frequencies are not equal (here we assume
$\omega_z\leq\omega_y\leq\omega_x$), while $m$ denotes the mass of
the corresponding particles.

The energy eigenvalues are
\begin{eqnarray}
E_{(n_x, n_y, n_z)} = \hbar\omega_x\Bigl[n_x + \frac{1}{2}\Bigr] +
\hbar\omega_y\Bigl[n_y + \frac{1}{2}\Bigr]
\nonumber \\
+ \hbar\omega_z\Bigl[n_z + \frac{1}{2}\Bigr] -
\frac{1}{2}\frac{mg^2}{\omega_z^3}\label{eneigen1}.
\end{eqnarray}

The density of states is calculated in the usual way
\cite{Phatria, Pethick}, namely, we resort to the continuum
approximation and consider the number of states proportional to
the volume of the surface in the first octant bounded by plane $E
= \frac{m}{2}\Bigl[\omega_x^2x^2 + \omega_y^2y^2 +
\omega_z^2z^2\Bigr]  + mgz$. The derivative, with respect to $E$,
provides the density of states (number of states per energy unit).
\begin{eqnarray}
\Omega(E)= \frac{[E + \frac{mg^2}{2\hbar\omega^2_z}-
\frac{\hbar}{2}(\omega_x + \omega_y +
\omega_z)]^2}{2\hbar^3\omega_x\omega_y\omega_z} \label{eneigen2}.
\end{eqnarray}

The density of the gas (number of particles per volume unit) is
given by (\ref{density1}). At this point we face the following
question: near the condensation temperature, $T_c$, how many
excited state wave functions shall be considered in
(\ref{density1})? As a rough approximation we will consider first
only one excited state. In addition, according to the bosonic
statistics, below the condensation tempe\-ra\-ture the average
number of particles in the ground state reads (for the case in
which the confining potential is an anisotropic harmonic
oscillator \cite{Pethick})
\begin{eqnarray}
 \langle n_{(0)}\rangle   = N\Bigl[1-\Bigl(\frac{T}{T_c}\Bigr)^3\Bigr]
 \label{occnumber12}.
\end{eqnarray}

Additionally, the average number of particles in the first excited
state is
\begin{eqnarray}
 \langle n_{(1)}\rangle   = N\Bigl(\frac{T}{T_c}\Bigr)^3
 \label{occnumber11}.
\end{eqnarray}

Therefore, under these conditions we have that
\begin{eqnarray}
 n(\vec{r}) = N\vert\phi_{(0)}(\vec{r})\vert^2\Bigl\{1
 +\Bigl(\frac{T}{T_c}\Bigr)^3\Bigl[\frac{2m\omega_z}{\hbar}(z +
 \frac{g}{\omega^2_z})^2- 1\Bigr]\Bigr\} \label{density2} .
\end{eqnarray}

In these last expressions $N$ denotes the number of particles
comprising the atom cloud and $\phi_{(0)}(\vec{r})$ is the ground
state wave function given as follows
\begin{eqnarray}
\vert\phi_{(0)}(\vec{r})\vert^2 =
\Bigl(\frac{m}{\pi\hbar}\Bigr)^{3/2}
 \Bigl(\omega_x\omega_y\omega_z\Bigr)^{1/2} \nonumber\\
\times \exp\Bigl\{-\frac{m}{\hbar}\Bigl[\omega_xx^2 +\omega_yy^2 +
\omega_z(z +
 \frac{g}{\omega^2_z})^2\Bigr]\Bigr\} \label{gswavefunction} .
\end{eqnarray}

The relation between the refraction index $\tilde{n}(\vec{r})$ and
the density $n(\vec{r})$ is provided by (\ref{refraction1}) (see
\cite{Newton} pp. 26) and it takes the form
\begin{eqnarray}
 \tilde{n}(\vec{r})= 1 + \lambda^2RN\vert\phi_{(0)}(\vec{r})\vert^2 \nonumber \\
 \times \Bigl\{1+\Bigl(\frac{T}{T_c}\Bigr)^3\Bigl[\frac{2m\omega_z}{\hbar}(z
 + \frac{g}{\omega^2_z})^2- 1\Bigr]\Bigr\}\label{refraction5} .
\end{eqnarray}

The arrival at the nonlinear material of our beams will show a
difference in time due to the fact that the reference beam has
always a speed equal to $c$, whereas the second beam will have a
lower speed, $v$, during its pass through the condensate. The
refraction index is defined as $\tilde{n}(\vec{r}) = c/v$
\cite{Guenther1}. The velocity inside the condensate is
position--dependent, i.e.,
\begin{eqnarray}
 v = c\Bigl\{1 + \lambda^2RN\vert\phi_{(0)}(\vec{r})\vert^2 \nonumber\\
 \Bigl(1+ \bigl(\frac{T}{T_c}\bigr)^3\Bigl[2\frac{m\omega_z}
{\hbar}(z + \frac{g}{\omega^2_z})^2-1\Bigr]\Bigr)\Bigr\}^{-1}
\label{velocity1} .
\end{eqnarray}

For the sake of clarity let us assume that the second beam during
its movement inside the condensate has coordinates $z$ and $y$
constant. The calculation of the difference in optical path
requires the knowledge of the distance that the light beam travels
inside the condensate. At this point we introduce a distance
parameter stating that the size of the condensate along the
$x$--axis is fixed by the harmonic oscillator length (see
\cite{Dalfovo} pp. 467) $ l^2_x = \hbar/m\omega_x$.

If $x_i$ and $x_f$ are the coordinates defining this width, then
$2l_x = x_f-x_i$. Let us denote by $\Delta t$ the time required by
the second beam to move from $x_i$ to $x_f$. Then
\begin{eqnarray}
 \Delta t = \int_{x_i}^{x_f}dx/v \label{time11} .
\end{eqnarray}

Explicitly,
\begin{eqnarray}
 \Delta t = \frac{2l_x}{c} +
 \frac{\lambda^2RIN}{c}
\alpha(z,y)\Bigl[1 + \beta(z,y)\bigl(\frac{T}{T_c}\bigr)^3 \Bigr]
\label{timeflight1}.
\end{eqnarray}

Define now $\omega_p = \Bigl(\omega_x\omega_y\omega_z\Bigr)^{1/3}$
and $\gamma = \Bigl(\frac{m\omega_p}{\pi\hbar}\Bigr)^{3/2}$ then
\begin{eqnarray}
\alpha(z, y)=
\gamma\exp\Bigl\{-\frac{m}{\hbar}\Bigl[\omega_yy^2+\omega_z(z +
\frac{g}{\omega^2_z})^2\Bigr]\Bigr\}\label{Def11},
\end{eqnarray}

\begin{eqnarray}
\beta(z, y)= \Bigl[2\frac{m\omega_z}{\hbar}(z +
\frac{g}{\omega^2_z})^2 -1\Bigr]\label{Def12},
\end{eqnarray}

\begin{eqnarray}
I
=\int_{x_i}^{x_f}\exp\{-\frac{m\omega_x}{\hbar}x^2\}dx\label{Def1},
\end{eqnarray}

\begin{eqnarray}
L = \Bigl(\frac{\hbar}{m\omega_p}\Bigr)^{1/2}\, , \quad l_i =
\Bigl(\frac{\hbar}{m\omega_i}\Bigr)^{1/2},~i = x, y,
z.\label{Def3}
\end{eqnarray}

Expression (\ref{timeflight1}) is the main result of the present
work. Notice that it contains a temperature--dependence of the
time--of--flight inside the condensate. This is the point to be
exploited. It is noteworthy to comment that Maxwell--Boltzmann
statistics has not been used at all.

%********************************************
%\subsection{Experimental feasibility}

Our experimental proposal is the following one: A ultra--intense
light pulse is emitted and divided into two parts. The upper beam
travels through a rubidium condensate with a certain value of $z$,
say $z_1$, we also keep its $y$--coordinate constant. i.e.,
$y=y_1$. We choose rubidium since its refraction index has already
been determined by interferometric methods, i.e., the technology
that the present proposal requires is nowadays available
\cite{Libbrecht}. Afterwards, both beams are brought together and
they impinge upon a nonlinear optical material. The difference in
optical lengths between the paths of our two beams entails that in
general they will not be in phase upon their arrival at the
nonlinear medium. The measurement of temperature will be done
using the characteristics behind the appearance of nonlinear
response. Indeed, the generation of high harmonics in a nonlinear
material depends upon several conditions, and one of them is
related to a threshold in the intensity (denoted here by $I_0$) of
the corresponding light pulse \cite{Franken}. If the beams are out
of phase, then the total intensity of the combined beam will be
smaller than $I_0$ \cite{Guenther1} and, in consequence, the
difference in time--of--flight of the two beams implies the
absence of high harmonics in the response of the nonlinear
material. If we recover the nonlinear response, then we may assert
that the beams are in phase. To achieve this we now modify the
length of the upper arm of the interferometer until high harmonics
emerge. In this sense we know that the difference in optical
length is determined by two variables: (i) the difference in
length between the upper and lower arms; (ii) presence of the
condensate. With the presence of high harmonics we may state that
the difference in optical length equals a multiple of the
wavelength of the corresponding light. This allows us to
determine, experimentally, the difference in optical length
induced by the condensate. The precision is provided only by
nonlinear optics, whereas in the usual model it involves not only
the precision associated to the imaging method \cite{Davis1} but
also the error introduced by the use of classical statistics.

We now change the length of the upper arm in such a way that at an
increase of it equal to $\Delta l$ the nonlinearity appears once
again. This last remark implies \cite{Chartier}
\begin{eqnarray}
\Delta l+ c\Delta t = \lambda.\label{difflength1}
\end{eqnarray}

Quantities $\Delta l$ and  $\lambda$ are detected experimentally,
while with the help of (\ref{timeflight1}) we determine $\Delta
t$. This provides the temperature
\begin{eqnarray}
\Bigl(\frac{T}{T_c}\Bigr)^3 =
\frac{1}{\beta(z_1,y_1)}\Bigl[\frac{\tilde
l}{\alpha(z_1,y_1)\lambda^2RIN} -1\Bigr] \label{time14}.
\end{eqnarray}

Where $\tilde l= \lambda -2l_x -\Delta l$. Let us now estimate the
feasibility of the idea. Assuming $N/L^3\sim 10^{15}cm^{-3}$ (see
\cite{Pethick} pp. 5), with $l_x\sim 10^{-1}cm$ \cite{Petrich},
considering a neodymium laser $\lambda \sim 10^{-4}cm$
\cite{Chartier}, and with $R$ equal to Bohr's radius, i.e., $R\sim
10^{-9}cm$, we obtain as estimation $\Delta l\sim 10^{-3}cm$. Our
argument also proves that the experimental values fulfill the
condition behind (\ref{refraction1}), namely,
$\lambda^2Rn(\vec{r})\sim 10^{-2}<1$ \cite{Newton}.

We now consider a more realistic situation, namely, below $T_c$
one finds that $q$ excited states are populated. Then
(\ref{density2}) becomes
\begin{eqnarray}
 n(\vec{r}) = N\vert\phi_{(0)}(\vec{r})\vert^2\Bigl\{1
 +\Bigl(\frac{T}{T_c}\Bigr)^3\nonumber\\
 \Bigl[\sum_{s=1}^q
 \frac{\exp\bigl\{-\tilde{E}_{s}/\kappa
 T\bigr\}}{2^ss!}H^2_s([z + g/\omega^2_z]/l_z)-1\Bigr]\Bigr\}
\label{density3} .
\end{eqnarray}

In this last expression $H_n(x)$ denotes the Hermite polynomials
\cite{Gradshteyn} and $\tilde{E}_{n} = \hbar\omega_zn$. We now
proceed as before. Indeed, with  (\ref{density3}) we obtain the
refraction index (resorting to (\ref{refraction1})) and calculate
the difference in time--of--flight. Let us now evaluate the order
of magnitude of the contributions stemming from those excited
states higher than the first one. In order to do this consider
$T=10^{-5}K$ \cite{Pethick} and also two excited states. Since
$H(x)_s\sim x^s$ (when $x>>1$) then $H^2_s(1 +
g/l_z\omega^2_z)\sim (1 + g/l_z\omega^2_z)^{2s}$. Therefore

\begin{eqnarray}
\sum_{s=1}^2
 \frac{\exp\bigl\{-\tilde{E}_{s}/\kappa
 T\bigr\}}{2^ss!}H^2_s([z + g/\omega^2_z]/l_z)\sim\nonumber\\
 \exp\bigl\{-\hbar\omega_z/\kappa T\bigr\}(1 + g/l_z\omega^2_z)^{2}
 \Bigl[1 + \nonumber\\
 \exp\bigl\{-\hbar\omega_z/\kappa T\bigr\}(1 + g/l_z\omega^2_z)^{2}\Bigr]\label{ordemag1}.
\end{eqnarray}

Introducing into (\ref{ordemag1}) the previous experimental values
we obtain that the contribution of the second excited state is
several orders of magnitude smaller than contribution of the first
excited state. This argument can easily be generalized to $q\geq3$
excited states and it proves that for rubidium at $T=10^{-5}K$
only one excited state suffices for deducing the required time--of
flight.

Without the use of Maxwell--Boltzmann statistics temperature could
also be detected by means of an interferometric procedure, but
this method has the shortcoming of requiring densities which could
imply problems to the current technology. This last remark is not
the proposal contained in \cite{Davis1} in which temperature was
measured with interferometry and using Maxwell--Boltzmann
statistics.

Let us mention that our idea could allow us to put forward a way
in which some postulates behind general relativity could be tested
\cite{Will}, i.e., our expressions depend upon $g$. Remember that,
up to now, there is no consistent quantum gravity theory, and that
any proposal shedding light upon the validity of the postulates
behind general relativity at quantal realm could be relevant
\cite{Time}.

%\begin{acknowledgments}
We would like to thank J. L. Hern\'andez--Pozos and R. Walser for
the fruitful discussions and literature hints. This research was
partially supported by CONACyT grants 47000--F, 48404--F, and by
the M\'exico-Germany collaboration grants CONACyT--DFG
J110.491/2006 and J110.492/2006. LFBG was supported by CONACyT
Grant No. 208211.
%\end{acknowledgments}

\end{document}